\documentclass[aps,prl,twocolumn,showpacs,superscriptaddress,floatfix]{revtex4-1}
\usepackage{amsmath,amssymb}
\usepackage{graphicx}
\usepackage{epsfig}
\usepackage{color}
\usepackage{rotating}
\usepackage{bm}
\usepackage{setspace}

\begin{document}
\title{Experimental Signature of Topological Superconductivity and Majorana Zero Modes on $\beta$-Bi$_2$Pd Thin Films}
\author{Yan-Feng Lv}
\author{Wen-Lin Wang}
\author{Yi-Min Zhang}
\author{Hao Ding}
\affiliation{State Key Laboratory of Low-Dimensional Quantum Physics, Department of Physics, Tsinghua University, Beijing 100084, China}
\author{Wei Li}
\author{Lili Wang}
\author{Ke He}
\author{Can-Li Song}
\email[]{clsong07@mail.tsinghua.edu.cn}
\author{Xu-Cun Ma}
\author{Qi-Kun Xue}
\email[]{qkxue@mail.tsinghua.edu.cn}
\affiliation{State Key Laboratory of Low-Dimensional Quantum Physics, Department of Physics, Tsinghua University, Beijing 100084, China}
\affiliation{Collaborative Innovation Center of Quantum Matter, Beijing 100084, China}
\date{\today}
\begin{abstract}
Using a cryogenic scanning tunneling microscopy, we report the signature of topologically nontrivial superconductivity on a single material of $\beta$-Bi$_2$Pd films grown by molecular beam epitaxy. The superconducting gap associated with spinless odd-parity pairing opens on the surface and appears much larger than the bulk one due to the Dirac-fermion enhanced parity mixing of surface pair potential. Zero bias conductance peaks, probably from Majorana zero modes (MZMs) supported by such superconducting states, are identified at magnetic vortices. The superconductivity exhibits resistance to nonmagnetic defects, characteristic of time-reversal-invariant topological superconductors. Our study reveals $\beta$-Bi$_2$Pd as a prime candidate for topological superconductor.
\end{abstract}
\pacs{74.55.+v, 68.65.-k, 74.25.Ha, 74.25.Jb}
\maketitle
\begin{spacing}{1.007}
Topological superconductors (TSCs) are a novel quantum phase of matter characterized by a fully gapped bulk state and gapless boundary states hosting exotic Majorana fermions that are their own anti-particles \cite{majorana1937symmetric}. The Majorana fermions obey non-Abelian braiding statistics and could be useful for fault-tolerant quantum computors \cite{kitaev2001unpaired, Elliott2015majorana}. Following theoretical proposals\cite{Fu2008superconducting, Qi2009time, Sau2010generic, Oreg2010helical, Choy2011majorana}, several experiments have disclosed their signatures in semiconductor nanowires \cite{mourik2012signatures, albrecht2016exponential}, iron atomic chains \cite{nadj2014observation} and topological insulators \cite{Xu2015experimental,Sun2016majorana} by proximity to superconductors, all sharing complex hybrid heterostructures. Alternatively, the newly discovered single-component superconductors, such as Cu/Sr/Nb-doped Bi$_2$Se$_3$ \cite{Fu2010odd, Sasaki2011topological, liu2015superconductivity}, In-doped SnTe \cite{Novak2013unusual} and PbTaSe$_2$ \cite{guan2016superconducting}, have  been suggested as potential TSC candidates, but far from a final conclusion \cite{Levy2013experimental}.

Tetragonal Bi$_2$Pd (hereafter, $\beta$-Bi$_2$Pd) crystallizes into a simple CuZr$_2$-type (I4/mmm) structure [Fig.\ 1(a)], and exhibits classical $s$-wave bulk superconductivity with a transition temperature ($T_\textrm{c}$) close to 5.4 K \cite{imai2012superconductivity}. Intriguingly, it was recently demonstrated from angle-resolved photoemission spectroscopy (ARPES) that $\beta$-Bi$_2$Pd holds several topologically protected surface bands cross the Fermi level ($E_F$) \cite{sakano2015topologically}. The nontrivial surface states of $\beta$-Bi$_2$Pd are subject to a classical $s$-wave bulk pairing, which naturally satisfies the key ingredients of proximity-induced two-dimensional (2D) topological superconductivity near the surface \cite{Fu2008superconducting}. Here the proximity-induced electron pairing on the spin-momentum-locked topological surface has a nontrivial topology and is obliged to be effectively spinless $p$-wave so as to guarantee the pair wave function antisymmetric \cite{Fu2008superconducting, Hao2011surface, Mizushima2014dirac}. Such superconducting states are anticipated to carry Majorana zero mode (MZMs) at the end of magnetic vortex lines, and thus reignite numerous research interests in $\beta$-Bi$_2$Pd. However, the subsequent studies consistently reveal a conventional $s$-wave superconductivity \cite{kacmarcik2016single, Biswas2016fully, Che2016absence} and no MZM at vortices of $\beta$-Bi$_2$Pd single crystals \cite{Herrera2015magnetic}. In this work, we used a state-of-the-art molecular beam epitaxy (MBE) in ultrahigh vacuum (UHV) to prepare $\beta$-Bi$_2$Pd thin films on SrTiO$_3$(001) substrate and characterized their superconductivity with \textit{in situ} scanning tunneling microscope (STM). We found the experimental evidence for nontrivial and impurity-resistant superconducting gap opening on the surface, as well as possible MZMs at vortices.

Our experiments were conducted in a cryogenic STM apparatus, connected with a MBE chamber for sample preparation. The base pressure of both systems is better than $10^{-10}$ Torr. Nb-doped (0.05 wt\%) SrTiO$_3$(001) substrates were outgassed in UHV and then annealed at $1200^{\circ}$C to obtain clean surface. The $\beta$-Bi$_2$Pd films were prepared by co-evaporating high-purity Pd (99.99\%) and Bi (99.999\%) sources from standard Knudsen cells under Bi-rich condition, with the substrate held at 300 $\sim$ $350^{\circ}$C. More details are discussed in the the Supplemental Material \cite{supplementary}. Polycrystalline PtIr tip was cleaned in UHV and calibrated on MBE-grown Ag films prior to all STM measurements at 0.4 K, unless otherwise specified. The differential conductance $dI/dV$ spectra and maps were collected using a standard lock-in technique with a small bias modulation of 0.1 mV at 913 Hz.
\end{spacing}

\begin{figure*}[t]
\includegraphics[width=1.98\columnwidth]{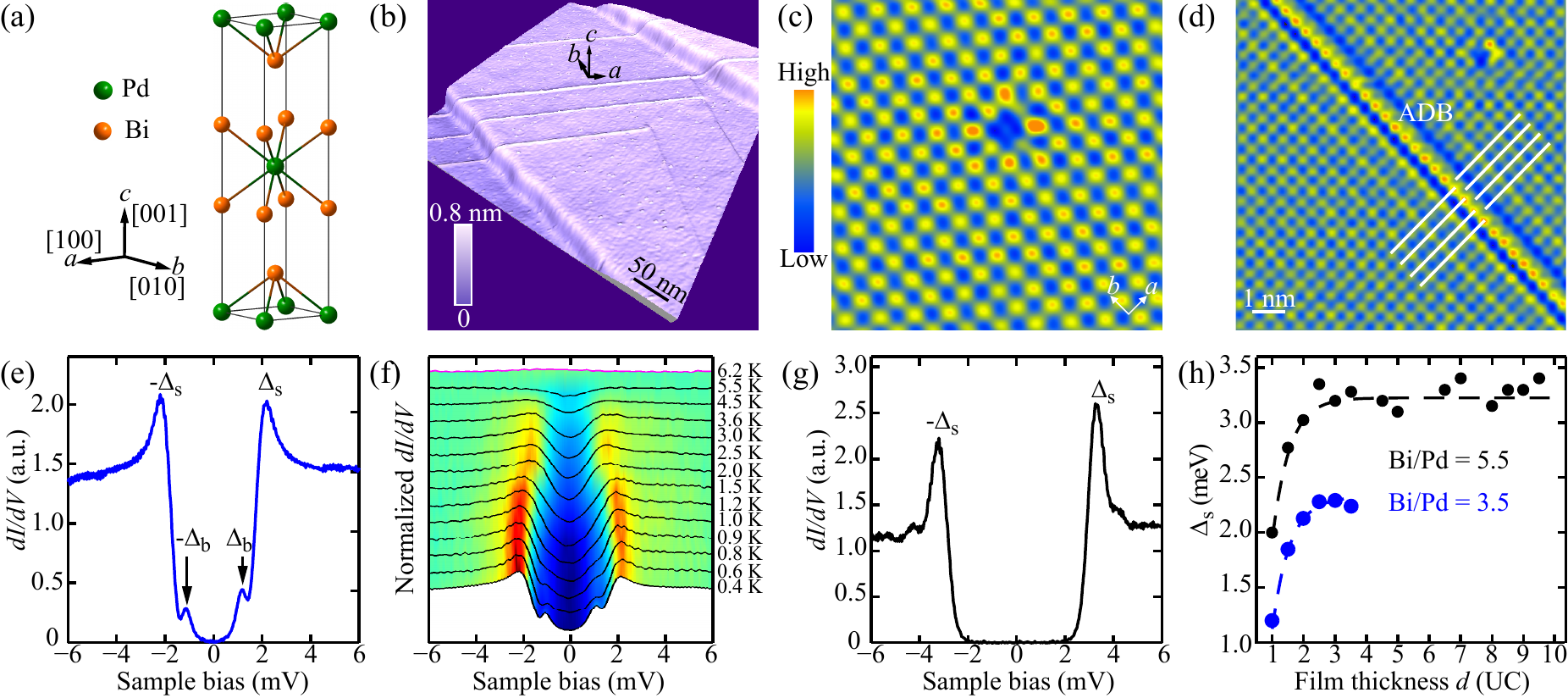}
\caption{(color online) (a) Crystal structure of $\beta$-Bi$_2$Pd. Each Pd atom sits at the center of the square prism of eight Bi atoms and the resulting Bi$_2$Pd motifs are stacked alternately by weak van der Waals forces, forming a body-centered layered structure. The $a$, $b$ and $c$ axes are taken along the crystal orientations. Along the \textit{c}(001) direction, each unit cell consists of two Bi-Pd-Bi triple layers. (b) STM topography ($V$ = 4.0 V, $I$ = 30 pA, 400 nm $\times$ 400 nm) of $\sim$20 UC thick $\beta$-Bi$_2$Pd films, grown with a Bi/Pd flux ratio of 3.1. The bright spots and linear protrusions correspond to Bi adatoms and ADB defects, respectively. The ADBs are orientated along either of the two in-plane crystallographic axes. Unless otherwise specified, our measurements are performed on this films. (c) Zoom-in image of $\beta$-Bi$_2$Pd surface ($V$ = 2 mV, $I$ = 280 pA, 4.5 nm $\times$ 4.5 nm), displaying a single Bi vacancy. The bright spheres indicate Bi atoms in the top layer. (d) A magnified ADB ($V$ = -1 mV, $I$ = 300 pA, 8 nm $\times$ 8 nm). The eight white lines show a lattice shift in the $b-c$ atomic planes across the ADB by half of the lattice parameter along the \textit{a} axis (1.7 \AA). (e) Differential conductance \textit{dI/dV} spectrum at 0.4 K, revealing two distinct superconducting gaps, denoted as $\Delta_\textrm{s}$ and $\Delta_\textrm{b}$, respectively. Spectra are acquired with a sample bias voltage of $V$ = 10 mV and tunneling current of $I$ = 300 pA throughout. (f) Temperature dependence of \textit{dI/dV} spectra in $\beta$-Bi$_2$Pd, displaying completely vanishing gap at 6.2 K (magenta curve). (g) \textit{dI/dV} spectrum on $\beta$-Bi$_2$Pd films prepared with a larger Bi/Pd flux ratio of 5.5. (h) Pairing gap $\Delta_\textrm{s}$ versus both the film thickness \textit{d} and Bi/Pd flux ratio. Dashed lines are guide to the eye.
}
\end{figure*}

MBE growth of $\beta$-Bi$_2$Pd films on SrTiO$_3$ substrate proceeds in Volmer-Weber mode. Epitaxial islands down to a single unit-cell (UC, two Bi-Pd-Bi triple layers) thick with lateral size of several hundreds of nanometers could be prepared, as seen from Fig.\ S1. We have established the growth dynamics of high-quality $\beta$-Bi$_2$Pd crystalline films by MBE, as detailed in the Supplemental Material \cite{supplementary}. Figure 1(b) shows a constant-current STM topographic image on the atomically flat Bi-terminated (001) $\beta$-Bi$_2$Pd films, grown by using a Bi/Pd flux ratio of 3.1. The films are found to straddle continuously over the neighboring terraces, indicative of a \lq\lq carpetlike\rq\rq\ growth of $\beta$-Bi$_2$Pd across the underlying SrTiO$_3$ steps. The adjacent Bi atoms are spaced $\sim$3.4 ${\textrm{\AA}}$ apart [Fig.\ 1(c)], while the out-of-plane lattice constant is approximate to 1.3 nm [Fig.\ S1]. They match excellently with those in $\beta$-Bi$_2$Pd crystals \cite{imai2012superconductivity} and thin films on Bi/Si(111) \cite{denisov2017growth}, justifying the chemical identity of epitaxial films studied here as $\beta$-Bi$_2$Pd. Because other Pd-Bi intermediate compounds all exhibit sharply different crystal structures and lattice parameters from our experimental observations \cite{okamoto1994bi}. Furthermore, the similar electron band structure between the MBE-grown thin films and $\beta$-Bi$_2$Pd crystals, revealed by ARPES \cite{imai2012superconductivity, denisov2017growth}, unambiguously backups this claim. Finally we found no other spurious phase in all samples investigated and the lattice constants alter little with film thickness $d$. This indicates a negligibly small strain involved, as expected for quasi van der Waals epitaxy of layered $\beta$-Bi$_2$Pd on SrTiO$_3$. Two distinct kinds of surface defects, namely Bi vacancy [Fig.\ 1(c)] and Bi adatom [Fig.\ 1(b)], as well as anti-phase domain boundary (ADB) [Fig.\ 1(d)] are identified.

Scanning tunneling spectroscopy (STS) probes the local density of states (DOS) and can measure the superconducting gap at $E_F$. In order to minimize the possible strain effects, we measured the tunneling conductance \textit{dI/dV} spectrum on a thicker ($\sim$26 nm) $\beta$-Bi$_2$Pd films at the base temperature of 0.4 K [Fig.\ 1(e)]. In sharp contrast to single-gap superconductivity reported for bulk counterpart \cite{kacmarcik2016single, Biswas2016fully, Che2016absence, Herrera2015magnetic}, two pairs of conductance peaks at two different energy scales are noticed, indicating double superconducting gaps in the $\beta$-Bi$_2$Pd films. The smaller one with a low spectral weight is estimated to be 1.0 $\pm$ 0.1 meV, close to the reported values of 0.76 $\sim$ 0.92 meV in bulk $\beta$-Bi$_2$Pd \cite{kacmarcik2016single, Biswas2016fully, Che2016absence, Herrera2015magnetic}. This hints at its possible origin from the bulk states and we thus dub it as $\Delta_\textrm{b}$ for simplicity. On the other hand, the newly discovered gap, which we reveal below stems from the topological surface states of $\beta$-Bi$_2$Pd and is dubbed as $\Delta_\textrm{s}$, is more prominent and significantly enhanced in magnitude.

\begin{figure*}[t]
\includegraphics[width=2\columnwidth]{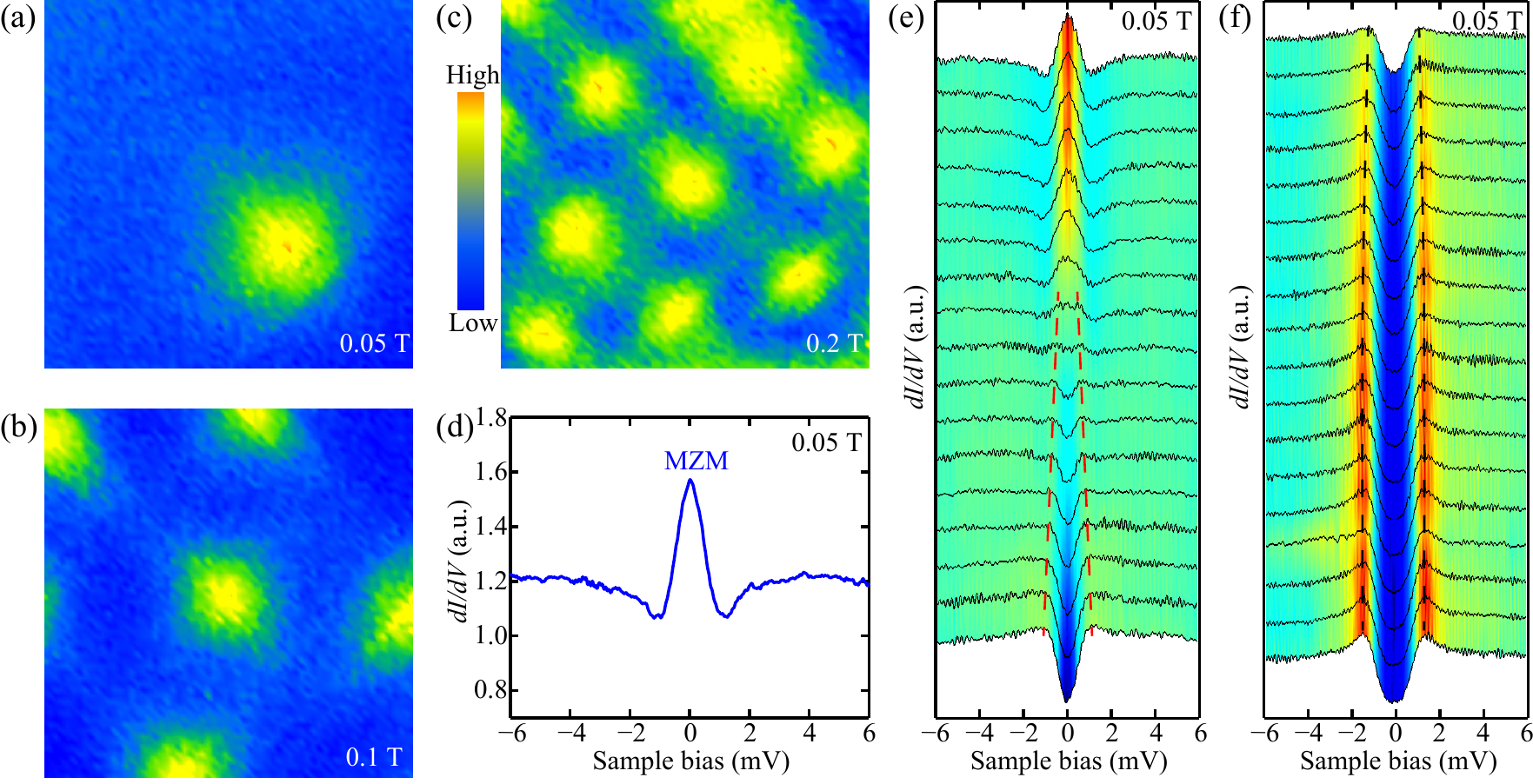}
\caption{(color online) (a-c) Normalized zero-bias conductance maps (setpoint: $V$ = 10 mV and $I$ = 200 pA, 350 nm $\times$ 350 nm) at varying magnetic fields, showing individual vortices. The normalization was performed by dividing all raw ZBC maps by the conductance value at a high sample bias of 6 meV. (d) Tunneling conductance \textit{dI/dV} spectrum taken at the vortex center, signifying the appearance of salient ZBCP. (e, f) Normalized \textit{dI/dV} spectra measured at locations with varying radial distance $r$ from the vortex center. Here $r$ alters from 0 nm (up) to 40 nm (down) with an equal separation of 2.5 nm in (e), and from 42 nm (up) to 144 nm (down) with an equal separation of 6 nm in (f), respectively. The ZBCPs are localized on a length scale of about 12 nm and show no splitting away from the vortex center. The dashed lines mark the evolution of $\Delta_\textrm{s}$.
}
\end{figure*}

The temperature-dependent \textit{dI/dV} spectra have been performed to decipher the origin of $\Delta_\textrm{s}$ in the epitaxial $\beta$-Bi$_2$Pd films, as plotted in Fig.\ 1(f). At elevated temperatures, the $\Delta_\textrm{s}$ is gradually smeared out and vanishes below 6.2 K. This result is surprising: the larger $\Delta_\textrm{s}$ of 2.15 meV would have led to a much higher $T_\textrm{c}$ (13 K or higher assuming the reduced gap ratio 2$\Delta/k_\textrm{B}T_\textrm{c} = 3.7\sim$ 4.1 for $\beta$-Bi$_2$Pd \cite{kacmarcik2016single, Biswas2016fully, Che2016absence, Herrera2015magnetic}) if it stems from another bulk band in $\beta$-Bi$_2$Pd \cite{imai2012superconductivity}. The other possibility is that $\Delta_\textrm{s}$ might has a root at the surface of epitaxial $\beta$-Bi$_2$Pd thin films, given the existence of topological surface states on $\beta$-Bi$_2$Pd \cite{sakano2015topologically}. In general, parity mixing of pair potential would occur near the surface of a superconductor due to the broken inversion symmetry there. Theoretically, the presence of Dirac fermions on the surface of $\beta$-Bi$_2$Pd \cite{sakano2015topologically} might anomalously enhance such mixing of $\Delta_\textrm{b}$ and time-reversal-symmetry (TRS) protected odd-parity pairing, leading to a larger pair potential $\Delta_\textrm{s}$ near the surface \cite{Mizushima2014dirac}. This appears a virtual match with our finding.

Note that the absence of $\Delta_\textrm{s}$ in $\beta$-Bi$_2$Pd single crystals \cite{Herrera2015magnetic}, as distinct from the epitaxial thin films, might be likely caused by a different $E_F$, where the surface states have merged into and are indistinguishable from the continuum bulk bands \cite{Hao2011surface, Mizushima2014dirac}. Theoretically, tuning the chemical potential to isolate the topological surface states from bulk bands near $E_F$ can bring about nontrivial electron pairing at the surface, as probably already realized here. A careful comparison of the electron states around $\Gamma$ does reveal an upward shift of $E_F$ ($\sim$70 meV) in $\beta$-Bi$_2$Pd/Bi/Si(111) thin films \cite{denisov2017growth} with respect to their bulk counterpart \cite{imai2012superconductivity}, although it is subject to a direct ARPES study of the $\beta$-Bi$_2$Pd films on SrTiO$_3$. Further defect engineering of $E_F$ by growing $\beta$-Bi$_2$Pd with a higher Bi/Pd flux ratio of 5.5, which may further increase the separation between the bulk and surface bands, leads to a larger $\Delta_\textrm{s}$ of 3.3 meV [Fig.\ 1(g)]. The enhanced surface gap $\Delta_\textrm{s}$ submerges $\Delta_\textrm{b}$ and has vanishing DOS over a finite energy range near $E_F$. This indicates a nodeless pairing gap function for $\Delta_\textrm{s}$, despite a slight anisotropy [Fig.\ S2]. Finally we find that $\Delta_\textrm{s}$ relies on the film thickness \textit{d} and reduces abruptly below $d \sim$2.5 UC [Figs.\ S2 and 1(h)]. This is probably resulted from the suppressed bulk superconductivity of $\beta$-Bi$_2$Pd when approaching the 2D limit. Since $\Delta_\textrm{s}$ links closely with $\Delta_\textrm{b}$ \cite{Mizushima2014dirac}, the reduced $\Delta_\textrm{b}$ will lead to a shrinking $\Delta_\textrm{s}$. The gradual saturation of $\Delta_\textrm{s}$ on thick films ($d \geq$ 3) echoes the above claim that the epitaxial stain is negligibly small in the $\beta$-Bi$_2$Pd films. As thus, we have revealed a possible TSC near the surface of epitaxial $\beta$-Bi$_2$Pd films.

In order to shed further light on $\Delta_\textrm{b}$, we have explored its dependence on magnetic field. Shown in Figs.\ 2(a-c) are the zero bias conductance (ZBC) maps in varying magnetic field, which was applied perpendicular to the sample surface. The yellow regions with enhanced ZBC due to the suppressed superconductivity signify individual isolated vortices, exhibiting a slight four-fold symmetry [Fig.\ S4]. This can be understood by the difference between coherence length $\xi$ along the crystal axes (\textit{a} or \textit{b}) and the diagonals, and matches with the observed gap anisotropy of $\Delta_\textrm{s}$ in Fig.\ S2 since $\xi \sim 1/\Delta$. Below the critical field $H_{\textrm{c2}}$, the vortex increases linearly in number with the field, as expected. Moreover, we found that the vortices tend to be preferentially pinned on ADBs (white dashes), leading to an overlapping of ADB-pinned magnetic vortices at the top-right corner of Fig.\ 2(c). Remarkably, each vortex core presents the \textit{dI/dV} spectrum with a salient ZBC peak (ZBCP) [Fig.\ 2(d)], which we will discuss later. Figures 2(e) and 2(f) depict the spatial dependence of \textit{dI/dV} spectra in the vicinity of single magnetic vortex. We notice that in stark contrast to usual superconductors the surface gap $\Delta_\textrm{s}$ does not recover to its zero-field value of 2.15 meV at position even larger than 100 nm from the vortex center [Fig.\ 2(f)]. This further supports the topologically nontrivial nature of $\Delta_\textrm{s}$, which we explain below.

\begin{figure}[b]
\includegraphics[width=\columnwidth]{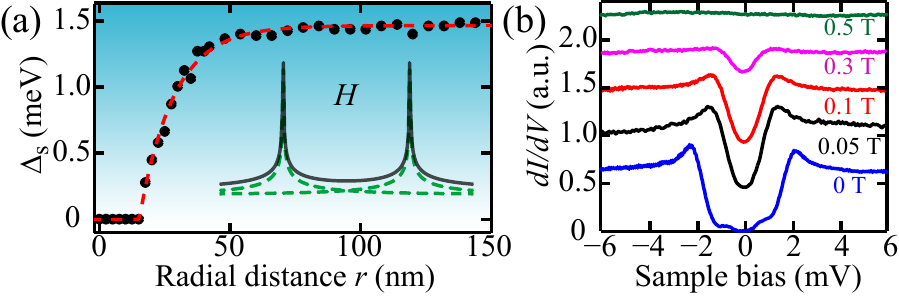}
\caption{(color online) (a) Gap magnitude of $\Delta_\textrm{s}$ plotted as a function of radial distance $r$ from the vortex center at 0.05 T. Note that $\Delta_\textrm{s}$ does not recover its zero-field value of 2.15 meV even far from the vortex center. Inset shows how the fields from isolated vortices (green dashes) overlap. Gray line corresponds to the overlapping field \textit{H}. (b) Field-dependent \textit{dI/dV} spectra in-between adjacent vortices, revealing a reduced gap $\Delta_\textrm{s}$ at elevated field. The zero-field spectrum (blue curve) is redrawn for comparison.
}
\end{figure}

\begin{figure}[b]
\includegraphics[width=\columnwidth]{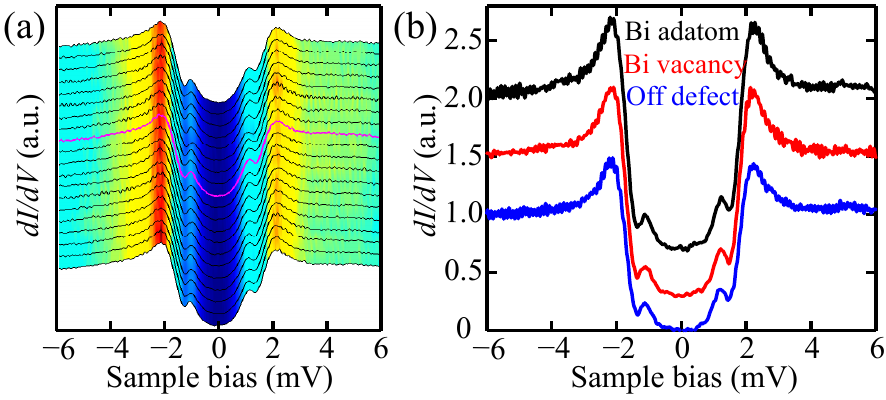}
\caption{(color online) (a) A series of \textit{dI/dV} spectra crossing an ADB, equally spaced 0.5 nm apart along a 10-nm trajectory. The magenta curve marks the spectrum on the ADB. (b) Comparison of superconducting energy gaps on and off Bi defects.
}
\end{figure}

Specifically, in common type-II superconductors the superconducting order parameter $\Delta$ regains its zero-field value outside the vortex core within the coherence length $\xi$, $\sim$23 nm in $\beta$-Bi$_2$Pd \cite{kacmarcik2016single}, while the magnetic field $H$ decays within a longer length scale (London penetration depth, $\lambda$). However, in TSCs, the magnetization around one vortex breaks the TRS and independently opens another insulating gap in the surface Dirac spectrum \cite{Fu2008superconducting, Burset2015superconducting}. This gap competes with and effectively weakens the superconducting gap $\Delta$. One thus anticipates that the superconductivity outside the vortex core of TSCs recovers in a longer length scale of $\lambda$ ($>$ 100 nm) other than $\xi$ \cite{kacmarcik2016single}. Our finding coincides with this expectation in a prominent manner. As summarized in Fig.\ 3(a), the surface gap $\Delta_\textrm{s}$ increases with the radial distance $r$ and seemingly saturates at 1.45 meV (still smaller than 2.15 meV) when $r$ exceeds three times of $\xi$ ($\sim$70 nm). Here the saturation behavior of $\Delta_\textrm{s}$ might be caused by the overlap of field \textit{H} from the circulating supercurrents of adjacent vortices, as schematically inserted in Fig.\ 3(a). At 0.05 T, the adjacent vortices are spaced only 219 nm. The fields around individual isolated vortices (green dashes) overlap, which consequently produces a very slowly varying \textit{H} in between (gray line). This leads to a negligibly small change of $\Delta_\textrm{s}$ with $r$ which is difficult to be resolved, when $r$ is moderately large. A further increase of the applied field enhances the overlapping field \textit{H} and consequently reduce $\Delta_\textrm{s}$ in between adjacent vortices. As shown in Fig.\ 3(b), our experiment indeed confirms this conjecture. It is worthy to note that this observation contrasts sharply from the behavior of trivial bulk gap $\Delta_\textrm{b}$ under magnetic fields, where the field \textit{H} fills the in-gap DOS but little changes the gap magnitude \cite{Herrera2015magnetic}. This observation further solidifies the topologically nontrivial nature of $\Delta_\textrm{s}$.

In distinction to the chiral $p$-wave superconductor, the effectively spinless superconductivity $\Delta_\textrm{s}$ realized here respects TRS and are topologically protected against scattering by TRS invariant nonmagnetic impurity \cite{Fu2008superconducting}, as the conventional \textit{s}-wave superconductor does \cite{anderson1959theory}. Figures 4(a) and 4(b) plot the \textit{dI/dV} spectra cross an ADB and on intrinsic Bi defects, respectively. No variation of $\Delta_\textrm{s}$ is observed as the STM tip is atop the ADB [Fig.\ 4(a)], Bi adatom and Bi vacancy [Fig.\ 4(b)]. This is consistent with the topological nature of $\Delta_\textrm{s}$ protected by TRS.

Finally we comment on the nature of ZBCPs identified at vortices [Fig.\ 2(d)]. Distinct from bulk $\beta$-Bi$_2$Pd crystals \cite{Herrera2015magnetic}, the ZBCPs observed here might originate from the ordinary Caroli-de Gennes-Matricon (CdGM) vortex bound states \cite{caroli1964bound} or MZMs \cite{Fu2008superconducting}. Since the topologically nontrivial superconducting states revealed above indeed harbor MZMs at vortices \cite{Fu2008superconducting}, it seems reasonable to ascribe the ZBCPs to MZMs. This is further supported by studying the spatial dependence of \textit{dI/dV} spectra in the vicinity of single magnetic vortex [Figs.\ 2(e) and 2(f)]. The ZBCP shows no splitting when moving away from the vortex center and is invariably fixed to the zero energy until the pairing gap $\Delta_\textrm{s}$ (marked by the dashed lines) begins to develop at a radial distance $r \sim$14 nm off the vortex core [Fig.\ 2(e)]. This differs markedly from the superconducting Bi$_2$Te$_3$/NbSe$_2$ heterostructure, where the MZMs entangle with usual CdGM bound states and result into a sophisticated space-dependent splitting of ZBCPs around the vortices \cite{Xu2015experimental,Sun2016majorana}. The identity of ZBCP observed as possible MZM (or at least contains the component of MZMs), rather than just trivial CdGM bound state, receives further evidence from the vanishing peak-dip structure outside $\Delta_\textrm{s}$ induced by the scattering states [Figs.\ 2(d) and 2(e)], which are often accompanied with trivial CdGM states bounded at vortices \cite{Gygi1991self}. Nevertheless, further investigation is desirable to fully understand the nature of ZBCPs observed here.

Our observation of TSC in one material of $\beta$-Bi$_2$Pd films has pointed to a novel avenue for searching topological pairing states on ordinary superconductors that possess topologically nontrivial surface states by $E_F$ engineering. This way dispels the structural complexity and interface unpredictability in artificial hybrid TSCs \cite{mourik2012signatures, albrecht2016exponential, nadj2014observation, Xu2015experimental, Sun2016majorana} and can mostly avoid the interruption from other non-topological alternatives (e.g.\ Kondo effect, disorder or impurity scattering) on MZMs \cite{Liu2012zero, Sau2015bound, Churchill2013superconductor}. As for $\beta$-Bi$_2$Pd, further growth control and electronic structure characterization are desired to finely tune its chemical potential and study the crossover from a topologically trivial superconductivity to the nontrivial one as reported here. Finally, analogous to the hybrid TSCs \cite{nadj2014observation, Xu2015experimental, Sun2016majorana}, the possible MZMs [Fig.\ 2(d)] are found to manifest a large peak width of $\sim$0.8 meV beyond the thermal broading, probably caused by their coupling to vortex-induced subgap states \cite{Das2016how}. An important next step in clarifying this issue would be to explore how such feimionic states affects the MZM broading.

\begin{acknowledgments}
We thank Y. J. Sun for helpful conversations. This work was financially supported by National Science Foundation, Ministry of Science and Technology and Ministry of Education of China. C. L. S acknowledges supports from the National Thousand-Young-Talents Program and the Tsinghua University Initiative Scientific Research Program.
\end{acknowledgments}

%

\widetext
\newpage
\setcounter{equation}{0}
\setcounter{figure}{0}
\setcounter{table}{0}
\setcounter{page}{5}
\makeatletter
\renewcommand{\theequation}{S\arabic{equation}}
\renewcommand{\thefigure}{S\arabic{figure}}
\renewcommand{\bibnumfmt}[1]{[S#1]}
\renewcommand{\citenumfont}[1]{S#1}

\onecolumngrid
\Large
{\textbf{Supplemental Material for:} }
\begin{center}
\textbf{\large Experimental Signature of Topological Superconductivity and Majorana Zero Modes on $\beta$-Bi$_2$Pd Thin Films}
\end{center}
\small
\maketitle
\twocolumngrid
\noindent
\begin{spacing}{1.8}
\textbf{Section 1: MBE growth of $\beta$-Bi$_2$Pd on SrTiO$_3$}

We prepared $\beta$-Bi$_2$Pd thin films on SrTiO$_3$(001) substrate by using Bi-rich condition. This idea is mainly formulated from the self-regulating growth mode of high-quality GaAs \cite{xue1997scanning}, topological insulators Bi(Sb)$_2$Se(Te)$_3$ \cite{li2010intrinsic, song2010topological, Jiang2012Fermi} and iron-based superconductor FeSe films \cite{Song2011molecular}. By choosing an appropriate temperature criterion, i.e.\ the substrate temperature ($T_{\textrm{sub}}$) lies between the temperatures of Knudsen cells holding the two reactants, the stoichiometry of epitaxial thin film is self-regulating. Guided by such growth criterion, we attempt to prepare $\beta$-Bi$_2$Pd films on SrTiO$_3$(001) substrate. However, we find that $\beta$-Bi$_2$Pd nucleates only when $T_{\textrm{sub}}$ is lower than both temperatures of Bi ($T_{\textrm{Bi}}$) and Pd ($T_{\textrm{Pd}}$) sources, namely $T_{\textrm{sub}} < T_{\textrm{Bi}} < T_{\textrm{Pd}}$. A higher $T_{\textrm{sub}}$ ($>$ $400^{\circ}$C) leads to no growth of Bi$_2$Pd compound.

\begin{figure*}[t]
\includegraphics[width=1.5\columnwidth]{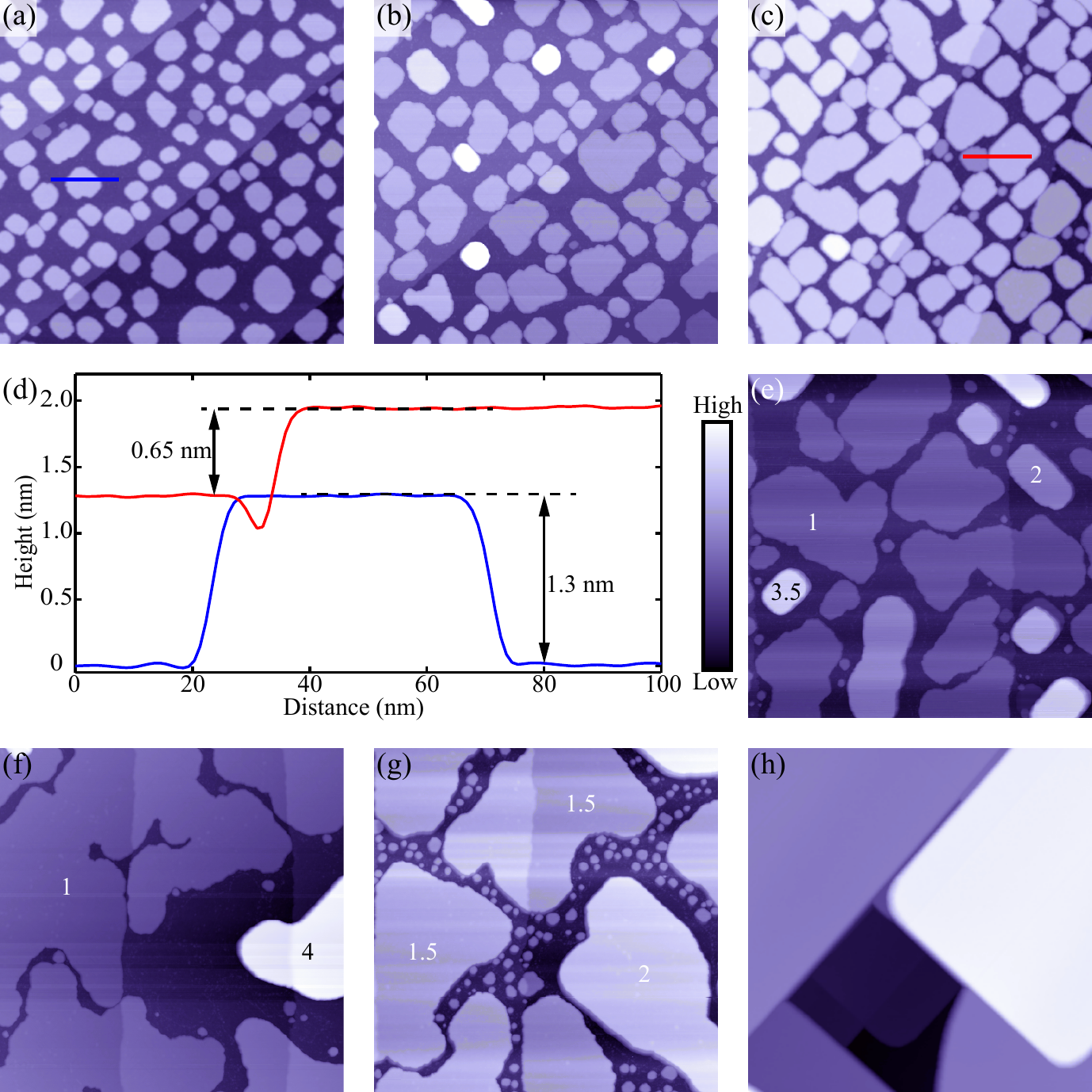}
\begin{spacing}{1.6}
\caption{MBE growth f $\beta$-Bi$_2$Pd films. (a-c) Topography evolution of $\beta$-Bi$_2$Pd films with increasing coverage with a Bi/Pd flux ratio of 5.5. (d) Profiles taken along the two colored lines in (a) and (c). Atomically flat terraces having a step height of $\sim$6.5 \AA suggest a half UC growth mode of $\beta$-Bi$_2$Pd on SrTiO$_3$(001). The minimum thickness of $\beta$-Bi$_2$Pd films is one unit-cell in (a). (e-h) STM topographies of coverage-dependent $\beta$-Bi$_2$Pd films with a Bi/Pd flux ratio of 3.5, showing fairly larger terraces. The numbers on the images denote the film thickness in terms of $\beta$-Bi$_2$Pd UC (13 \AA). Tunneling conditions and image size: $V$ = 3.5 V, $I$ = 30 pA, 500 nm $\times$ 500 nm except for (c) $V$ = 1.5 V, (e) $V$ = 5.0 V, and (h) 300 nm $\times$ 300 nm.}
\end{spacing}
\end{figure*}

In order to understand this behavior, we have studied the sticking coefficient $\tau$ of Bi on SrTiO$_3$ at various $T_{\textrm{sub}}$. It is found that $\tau$ is significantly small at $T_{\textrm{sub}} > $ $200^{\circ}$C. Therefore, as an appropriately higher $T_{\textrm{sub}}$ ($\sim$$300-350^{\circ}$C in our experiments) is chosen, the Bi behaves as volatile as As, Se and Te, and the above temperature criterion is effectively satisfied. Under these conditions, the extra Bi will not react with Pd to form $\beta$-Bi$_2$Pd unless an excessive Pd exists on the substrate. The growth of $\beta$-Bi$_2$Pd films has a ratio scaling linearly with Pd flux determined by $T_{\textrm{Pd}}$ and obeys a similar growth dynamics of GaAs, Bi(Sb)$_2$Se(Te)$_3$ and FeSe compounds \cite{xue1997scanning, li2010intrinsic, song2010topological, Jiang2012Fermi, Song2011molecular}. The only distinction is that a number of intrinsic defects always occur in the epitaxial $\beta$-Bi$_2$Pd films. We take advantage of this to engineer the Fermi level $E_F$ in $\beta$-Bi$_2$Pd. Given the Bi-rich condition, we speculate that the Bi-related defects dominate in number, dope electrons into $\beta$-Bi$_2$Pd films and move the $E_F$ upwards as compared to bulk $\beta$-Bi$_2$Pd crystals. This is because the electronegativity of Bi (2.02) is smaller than that (2.2) of Pd.

The optimal conditions for MBE growth of $\beta$-Bi$_2$Pd thin films are established by a systematic study of the STM topography under various growth parameters. Based on the criterion above, we find that the $\beta$-Bi$_2$Pd morphology depends critically on the Bi/Pd flux ratio. At a high Bi/Pd flux, many $\beta$-Bi$_2$Pd islands with very small lateral size \textit{L} nucleate, accompanied with unwanted clusters (not shown). As the Bi/Pd flux ratio is lowered to 5.5 [Fig.\ S1(a-c)], regular $\beta$-Bi$_2$Pd islands with intermediate \textit{L} are observed. Here we set $T_{\textrm{Pd}}$ = $1140^{\circ}$C and $T_{\textrm{Bi}}$ = $480-500^{\circ}$C, leading to a growth rate of $\sim$0.07 UC/min. When the Bi/Pd flux ratio is further reduced to 3.5 [Fig.\ S1(e-h)], $\beta$-Bi$_2$Pd islands with larger \textit{L} of hundreds of nanometers are obtained. The mechanism behind this behavior deserves a further study and lies beyond the scope of the current study. Nevertheless, the size-controlled $\beta$-Bi$_2$Pd islands provide a unique platform to investigate their superconducting properties and MZMs. More remarkably, the topological superconducting gap $\Delta_\textrm{s}$ increases in magnitude with increasing Bi/Pd flux ratio, as discussed in the main text. Although the samples grown with a higher Bi/Pd flux ratio are more favorable for studying the topological properties of $\beta$-Bi$_2$Pd superconductor, their small lateral size \textit{L} hinders the imaging of vortices and MZMs, which requests an extremely large \textit{L}. Therefore, in order to investigate MZMs at vortex and keep consistency, we have studied $\beta$-Bi$_2$Pd islands grown with a smaller Bi/Pd flux ratio of 3.1, unless otherwise specified. This sacrifices the surface pairing gap magnitude $\Delta_\textrm{s}$ to some extent.

\textbf{Section 2: Anisotropic nodeless pairing gap $\Delta_\textrm{s}$}

\begin{figure}[t]
\includegraphics[width=\columnwidth]{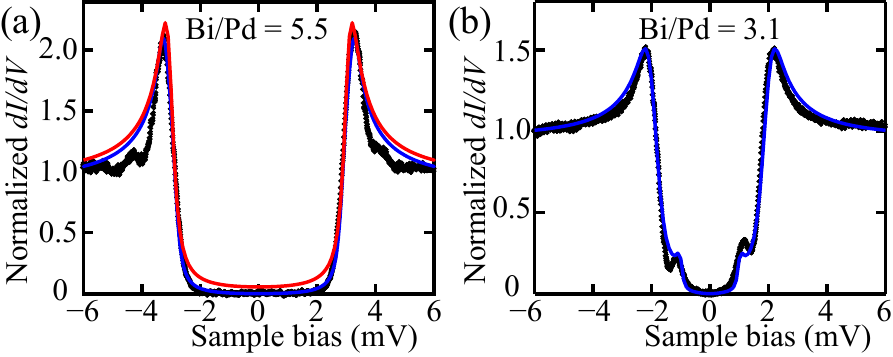}
\begin{spacing}{1.6}
\caption{Fits of $\Delta_\textrm{s}$ by an anisotropic nodeless pairing gap function. (a,b) Normalized \textit{dI/dV} spectra (black dots) on $\beta$-Bi$_2$Pd films grown with the Bi/Pd flux ratio of 5.5 [Fig.\ 1(g)] and 3.1 [Fig.\ 1(e)], respectively. The normalization is conducted by dividing the raw \textit{dI/dV} spectra by their backgrounds, extracted from a linear fit to the conductance outside the superconducting gap. The blue curves show the best fits of \textit{dI/dV} spectra to a single anisotropic gap $\Delta_\textrm{s}$ = 3.25 (0.94 $\pm$ 0.06$\cdot$cos4$\theta$) in a or a combination of an anisotropic gap $\Delta_\textrm{s}$ = 2.1 (0.93 $\pm$ 0.07$\cdot$cos4$\theta$) and an isotropic \textit{s}-wave gap $\Delta_\textrm{b}$ = 1.0 meV in (b), while the red curve in (a) to a single isotropic \textit{s}-wave gap of 3.03 meV.
}
\end{spacing}
\end{figure}

\begin{figure}[t]
\includegraphics[width=\columnwidth]{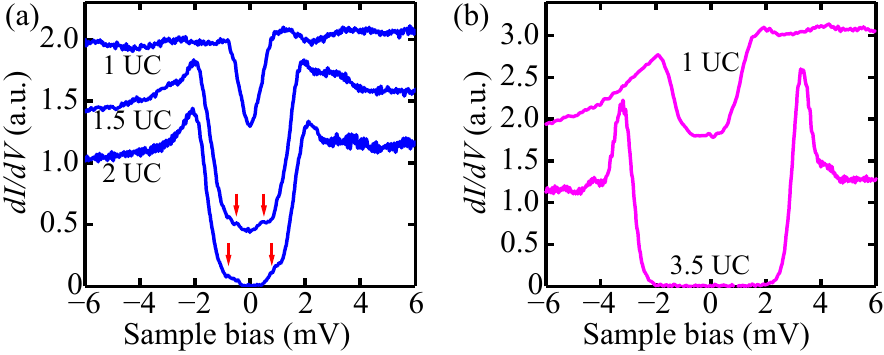}
\begin{spacing}{1.6}
\caption{Dependence of \textit{dI/dV} spectra on the film thickness \textit{d} at two different Bi/Pd flux ratio of (a) 3.5 and (b) 5.5. The gap magnitudes of $\Delta_\textrm{b}$ (red arrows) have been reduced to be 0.5 meV and 0.78 meV for 1.5 UC and 2.0 UC films in (a), respectively. Bulk gap $\Delta_\textrm{b}$ is indiscernible in (b), which might be primarily blurred by the significantly enhanced $\Delta_\textrm{s}$.
}
\end{spacing}
\end{figure}

\begin{figure}[t]
\includegraphics[width=0.75\columnwidth]{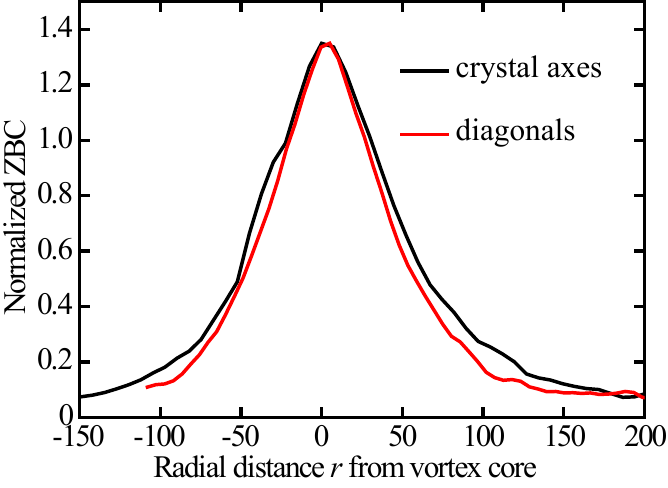}
\begin{spacing}{1.6}
\caption{Radial dependence of the normalized ZBC along the crystal axes (black line) and the diagonals (red line), respectively, indicating a slight anisotropy in vortex shape.
}
\end{spacing}
\end{figure}

Fits of the \textit{dI/dV} spectra with various gap functions shed mechanistic insight into the surface superconducting pairing gap $\Delta_\textrm{s}$. To do so, we choose two typical \textit{dI/dV} spectra in the main text, i.e.\ Fig.\ 1(e) and 1(g), which are redrawn in Fig.\ S2. Fits of more \textit{dI/dV} spectra lead to the same result. The \textit{dI/dV} spectra feature flat bottom with completely vanishing DOS over a finite energy range near $E_F$ and are U-shaped, suggesting nodeless pairing for $\Delta_\textrm{s}$. However, we find that the U-shaped spectra could never be satisfactorily fitted to an isotropic BCS DOS by the Dynes model \cite{Dynes1984tunneling}, as illustrated by the red curve in Fig.\ S2(a). Compared to an isotropic BCS gap, the relatively weak superconducting coherence peaks at $\pm\Delta_\textrm{s}$ (black dots) indicate that a small gap anisotropy may be involved. Indeed, anisotropic gap $\Delta_\textrm{s}$ = 3.25 (0.94 + 0.06$\cdot$cos4$\theta$) nicely follows the DOS within the measured superconducting gap $\Delta_\textrm{s}$ in Fig.\ 23(a) (blue curve). In Fig.\ S2(b), by adding a small fraction (8\%) of isotropic \textit{s}-wave gap responsible for $\Delta_\textrm{b}$, the \textit{dI/dV} data could also be nicely fitted to another anisotropic gap $\Delta_\textrm{s}$ = 2.1 (0.93 + 0.07$\cdot$cos4$\theta$). Here the electron-hole asymmetry of $\Delta_\textrm{b}$ in spectral weight might be caused by a sloped DOS in the normal state of $\beta$-Bi$_2$Pd superconductor. The nodeless superconducting pairing gap with a slight anisotropy places severe constraints on the possible odd-parity pairing symmetry for $\Delta_\textrm{s}$.

\end{spacing}
%

\end{document}